# Discretely guided electromagnetic effective medium

K. Kempa, X. Wang, Z.F. Ren, and M.J. Naughton

Department of Physics, Boston College, Chestnut Hill, MA 02467

A material comprised of an array of subwavelength coaxial waveguides decomposes incident electromagnetic waves into spatially discrete wave components, propagates these components without frequency cut-off, and reassembles them on the far side of the material. The propagation of these wave components is fully controlled by the physical properties of the waveguides and their geometrical distribution in the array. This allows for an exceptional degree of control over the electromagnetic response of this effective medium, with numerous potential applications. With the development of nanoscale subwavelength coaxial waveguides, these applications (including metamaterial functionality) can be enabled in the visible frequency range.

Artificial electromagnetic propagation systems comprised of subwavelength elements have recently been proposed, studied and engineered to produce varied, non-naturally-occurring combinations of effective electric permittivity $\varepsilon$ and magnetic permeability $\mu$, thus facilitating novel methods of controlling electromagnetic radiation [1-4]. It has been shown that such systems can produce various exotic effects, such as negative refraction [5], subwavelength lensing [6-8] and cloaking [9-12].

Recently, a *wire guided* effective medium, based on arrays of conducting wires, has been proposed, and shown to have a range of interesting properties, such as subwavelength, near field lensing [13,14]. The principle of operation of these systems is based on subwavelength wave guiding through the wire arrays (canalization), similar to that of a multicore coaxial wire [15]. While the electromagnetic response of this medium can be controlled by the array geometry, this control is limited by the *distributed* nature of the wave propagation in (and around) the array. Each wire can be considered as a waveguide of elementary waves, into which the incoming wave decomposes, but there is substantial crosstalk between these *unshielded* waveguides, severely limiting propagation control.

Here, we propose a *discretely guided medium* based on an array of *bona fide* (*i.e. shielded* center conductor) coaxial subwavelength waveguides piercing an otherwise nontransparent material. A schematic of this medium in various configurations is shown in Fig. 1. Fig. 1a represents a complex configuration with waveguides of different lengths, each line representing a coaxial waveguide. The dimensions of the waveguides transverse to the propagation direction, as well as the average inter-waveguide distance, are smaller than the radiation wavelength $\lambda$. Note that in this architecture, wave propagation occurs *only* via the coaxial waveguides, and not through the medium in between. The fundamental principle of operation of our discretely guided medium is based on Huygens' principle [16]: an arbitrary wave interacting with a medium can be Fourier-decomposed into plane waves. Each plane wave, in turn, excites the subwavelength waveguides of the medium. The excitations propagate along the waveguides, and are re-radiated on the other side of the medium. The resulting Huygens wavelets re-assemble into a plane wave, the propagation of which is controlled by the parameters of the waveguides (their distribution, length, direction, *etc*. inside the medium).

Each waveguide propagates radiation in a mode identical to, or closely resembling, an acoustic transverse electromagnetic (TEM) mode, familiar from the common coaxial cable [15,17,18] long used in radio technology. At higher frequencies, the dimensions of the coaxial waveguides must be reduced to accommodate the subwavelength limit, and so in the optical range, nanoscopically small coaxial waveguides (nanocoax) must be used. Such nanocoaxes have been recently fabricated [19], and shown [19-21] to propagate plasmon polaritons which, for frequencies sufficiently below the constituent metal plasma frequency(ies), closely resemble conventional TEM waveguide modes. Coupling of radiation to the nanocoax arrays is envisioned via antenna couplers. Recently, it has been demonstrated [22,23] that nanoantennas can process visible light much like conventional radio antennas process radio waves. The coupling efficiency can be calculated (or computer simulated) at radio frequencies, using standard radio engineering techniques [18]. The same is possible for visible frequencies, taking into account the complete dielectric response of the materials involved.

Consider a simple guided medium with subwavelength coaxial waveguides (and inter-waveguide medium) made of "good" metallic conductors (*i.e.* with low loss and large negative dielectric constant in the frequency of interest), and the inter-electrode space (inside the waveguides) filled with a low loss dielectric. These conditions, easy to assure in the radio frequency range, can also be achieved in the optical range. In Fig 1b, we assume that all coaxial waveguides are of equal length, and their ends form a square lattice on each side of the medium, with the same lattice constant. It is straightforward to conclude that any electromagnetic field pattern generated at one surface of this medium is transferred to other side with *subwavelength* resolution. This is because the transfer is achieved by *propagating* TEM-like modes of the subwavelength waveguides. Such modes do not experience frequency cut-off, and therefore the resolution will be limited *only* by the geometric parameters of the waveguide ends (diameter, areal density). This is in contrast to the case of a conventional translucent material, which is diffraction limited, since the large momentum Fourier components turn into evanescent waves.

Near-field lensing of the field intensity patterns projected onto the surface of the medium is enabled by employing the converging (or diverging) configuration depicted in Fig. 1c, *i.e.* with patterns of waveguide-ends having different lattice constants on each side of the medium. With nanoscopic waveguides, such a subwavelength lens may be used to increase resolution of conventional optical photolithography, or to improve near-field scanning optical microscopy (NSOM).

We next consider the far-field response of this medium, which can be computed, in general, for any distribution of waveguides and corresponding couplers on the emitting side. The problem is essentially that of a planar phased array [24,25]. For a system of radiators with inter-radiator distance much less than λ, the far-field response can be estimated. It can be shown [24] that if the lengths of the subwavelength waveguides $l_{mn}$ are chosen so that

$$l_{mn}\sqrt{\varepsilon_{mn}} = (\hat{n}_1 - \hat{n}_0) \cdot \vec{R}_{mn} + constant \qquad (1)$$

(where $\vec{R}_{mn}$ is the vector position of the (m,n) waveguide end in the square array and $\varepsilon_{mn}$ is its inter-electrode dielectric constant), the re-radiated far-field on the far-side, in response to the incoming plane wave in the $\hat{n}_0$ direction, is a simple plane wave propagating in a chosen direction $\hat{n}_1$. The significance of Eq. (1) lies in the realization that the propagation of an arbitrary incident wave through the medium can now be *fully* controlled.

According to Eq. (1), a medium with equal length coaxes, as in Fig. 1b, each filled with the same dielectric, assures that $\hat{n}_1 = \hat{n}_0$. In Fig. 2, we show this explicitly via simulation of a square array of straight, parallel nanocoaxes, employing a 3D finite-difference time-domain (FDTD) method to solve numerically Maxwell's equations (for details, see Ref. 26). The metal regions are chosen to be Al, described by the Drude dielectric function, and the inter-electrode dielectric is air. The instantaneous electric field magnitude around and inside the nanocoax structure, generated by an incoming plane wave along the direction shown, is encoded as a color map (red-to-blue is a change of sign). For clarity, only the field distribution in a small vertical strip through one lattice period is shown. Note that the wave emerging below the film is in the same direction as the incident, as expected. The transmission efficiency (~90%) is controlled by the geometry of the structure, including the length of the inner core protrusions (nanoantennas). The transmission is via a $TM_{00}$ mode, which is cut-off free, and which at low frequencies reduces to the TEM mode of the conventional coaxial waveguide [21].

The exceptional control over wave propagation offered by this guided medium can be used for various applications. The dielectric constant $\varepsilon_{mn}$ can be altered electronically (by employing piezoelectric dielectrics or active components inside the coaxial waveguide), controlling not only the propagation direction, but the wave polarization, phase, amplitude and frequency. Moreover, antenna-waveguide impedance mismatch and metallic losses along waveguides could be eliminated using amplifiers, mounted directly inside the waveguides. Discrete, electronic switching of light propagation can also be accomplished by "shorting" all, or a selected group, of waveguides, with similar active components. Non-linear optical components will allow for wave mixing and phase conjugation [27]. With nanoscopic elements, these could be possible in the visible range.

The discretely guided medium can simulate metamaterial effects as well. For example, electromagnetic "cloaking" of objects placed between the waveguides occurs. This is demonstrated in Fig. 1d, where the hole in the material between the waveguides forms an "exclusion zone". Since the antenna cross-sections can be made much larger than the physical area occupied by the waveguide-ends [18], there is a possibility, in-principle, to arrange waveguides so that the exclusion zone is large. Any object placed inside the zone will be invisible, as its presence does not alter the wave propagation, which occurs entirely through the coaxial waveguides. In Fig. 1e is the case of cross-wired waveguides, where entrance-exit waveguide-end configurations are image-reflected about the horizontal. Thus, this medium acts like a flat lens, in which intra-lens wave propagation is not diffraction limited. Note that the medium in this form is not capable of a true negative refraction, and therefore this is not a superlens in the Pendry sense [6]. However, since phase conjugation is possible in our medium, true negative refraction might be possible [28].

In conclusion, we propose the concept of a guided medium, based on subwavelength coaxial waveguides, which can be engineered to produce a wide range of desired electromagnetic responses. This guided medium decomposes incoming light waves into partial, propagating waveguide modes, which in turn re-assemble into propagating waves on the other side of the medium. Since the subwavelength waveguides do not experience frequency cut-off, a light wave is transmitted through the medium without loss of subwavelength resolution. Additional control over propagation in the waveguides can be obtained via active, non-linear nano-components. With nanoscopic, coaxial nanowaveguides, such as have recently been invented [19], operation of this medium can be extended to the visible frequency range.

**Figure Captions**

**Fig. 1.** Schematic of the guided medium with (a) waveguides of different lengths in regular arrays with equal lattice constants on both sides, (b) waveguides of equal lengths in regular arrays with equal lattice constants on both sides, (c) wave-guides of different lengths in regular arrays with different lattice constants on both sides, (d) example of a cloak, and (e) example of a subwavelength lens.

**Fig. 2** FDTD simulation of the transmission of light ($\lambda = 500$ nm) through a simple nanocoax medium. Film thickness is 800 nm, array period 400 nm, diameter of the inner core of the nanocoax 100 nm, and the inner diameter of the outer electrode 300 nm. At bottom is a top view.

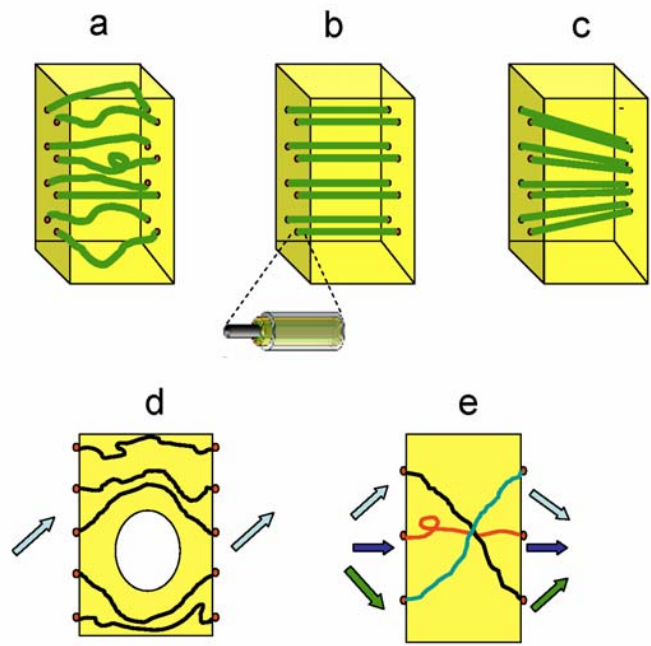

**Fig. 1**

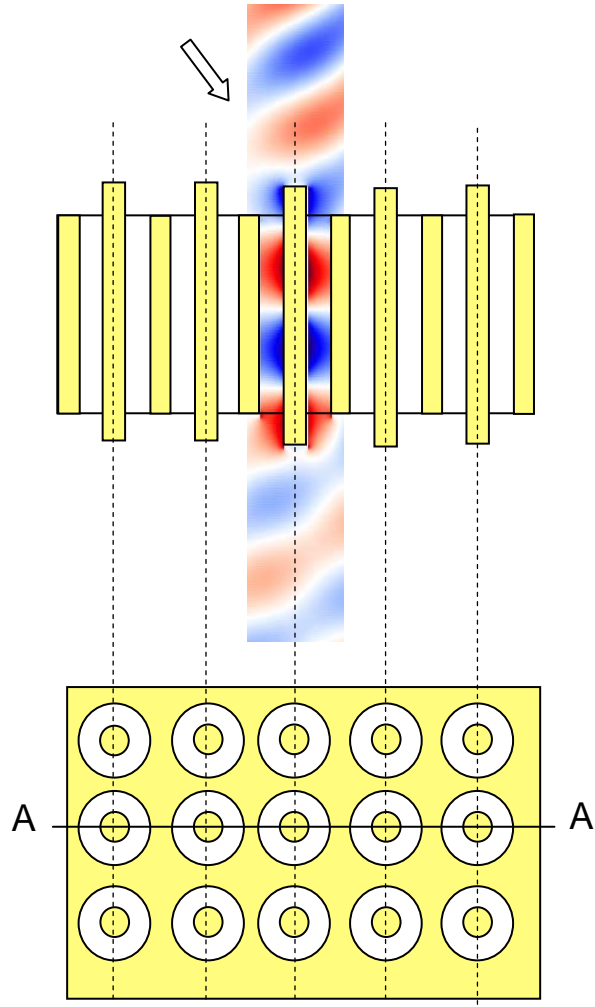

**Fig. 2**